# Spectroscopic investigation of phase transitions in 1-nonanol and 1-decanol

Yelyzaveta Chernolevska[a], Yevhenii Vaskivskyi[a], Valeriy Pogorelov[a],

Iryna Doroshenko[a], Olena Doroshenko[a], Valdas Sablinskas[b], Vytautas Balevicius[b]

[a] Taras Shevchenko National University of Kyiv, Volodymyrska str., 64/13, Kyiv, Ukraine
[b] Vilnius University, Sauletekio 9-3, Vilnius-10222, Lithuania

**Abstract**

FTIR spectra of monohydric alcohols 1-nonanol ($C_9H_{19}OH$) and 1-decanol ($C_{10}H_{21}OH$) were registered in the spectral region from 500 cm$^{-1}$ to 4000 cm$^{-1}$ at temperatures from -50 to + 25 °C for 1-nonanol and +100 °C for 1-decanol. Temperature-induced spectra changes were compared for these two alcohols. The authors link the observed changes with the transformations of cluster structure occurring during phase transitions.

## 1. Introduction

Investigations of partly ordered liquids with hydrogen bonding (alcohols, water, etc.) are of interest for physics, chemistry, biology and other nature sciences. Alcohols are important for industrial and technical applications: they are often used as solvents; are a part of surfactants; can be used for polymer materials synthesis. Alcohols are also used in medicine and food industry. In view of the above, there is no field, where these substances are not used.

Molecules of alcohols have a possibility to form various structures (clusters) due to hydrogen bonding [1 - 3]. For small alcohols (such as methanol or ethanol), there is no spectral band in the region of free O-H stretching vibrations (3600 – 3700 cm$^{-1}$) in condensed phase [4 - 6]. In literature, different authors demonstrate experimental FTIR spectra [1, 7 - 10] with bands in the region about 3200 cm$^{-1}$. In support of this, from low-temperature experiments in matrix isolation it is seen, that vibrational bands at 3200 – 3400 cm$^{-1}$ appear in spectra of alcohols during heating of the sample [11 - 14]. At the same time, bands in the region 3600 – 3700 cm$^{-1}$ disappear. After matrix heating spectra become similar to spectra in condensed phase [11 – 14]. All these changes can be interpreted as restructuring of cluster structures in alcohols.

Most researchers have focused on investigations of the simplest alcohols because there is a possibility to apply both experimental and theoretical techniques to such kinds of investigations [1, 15 – 19], but articles about bigger alcohols cannot be found at all. In this work, we present new experimental results on alcohols with bigger molecules - 1-nonanol (fig. 1) and 1-decanol (fig. 2).

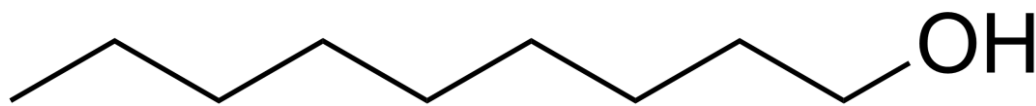

Fig. 1. Skeletal formula of 1-nonanol

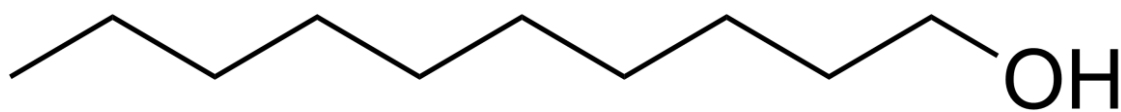

Fig. 2. Skeletal formula of 1-decanol

## 2. Experimental details

All experimental spectra presented in our article were registered in the laboratory of Fourier transform infrared absorption spectroscopy at Vilnius University (Lithuania). Liquid 1-nonanol and 1-decanol with purity > 99.9 % from Fluka were used as received. The spectra were registered by attenuated total reflection (ATR) method using FTIR spectrometer VERTEX 70 from Bruker in the spectral region from 500 to 4000 cm$^{-1}$. Single-pass ZnSe ATR crystal was used for capturing of the ATR spectra. Angle of incidence of IR beam was set to 70 degrees, what insured total reflection from ZnSe/alcohol interface. Spectral resolution was set to 1 cm$^{-1}$ and in order to increase signal-to-noise ratio each spectrum was taken as an average of 64 scans. The spectrometer was equipped with a liquid-N$_2$-cooled mercury cadmium telluride (MCT) detector. Blackman-Harris 3-term apodization function and zero filling 2 were used during processing of interferograms. Temperature investigations were carried out in temperature range from -50 to + 25 °C for 1-nonanol and +100 °C 1-decanol by gradual heating. LINKAM cryostat (model FTIR 600) was used for thermostabilization.

## 3. Results and discussion

In fig. 3 and 4, temperature-induced evolutions of the registered FTIR spectra of 1-nonanol and 1-decanol in the spectral region 3000 – 3700 cm$^{-1}$ are presented. This is the spectral region of O-H stretching vibrations. Here we cannot see drastic changes near the phase transition solid-liquid. For 1-nonanol (fig. 3), the maximum position smoothly shifts towards high frequency region upon temperature increasing. On the other hand, for 1-decanol (fig. 4), the positions of two maxima stay at the same places up to the phase transition, then "jump" as one single peak to the low frequency region just after the phase transition and after that start to shift towards higher frequencies.

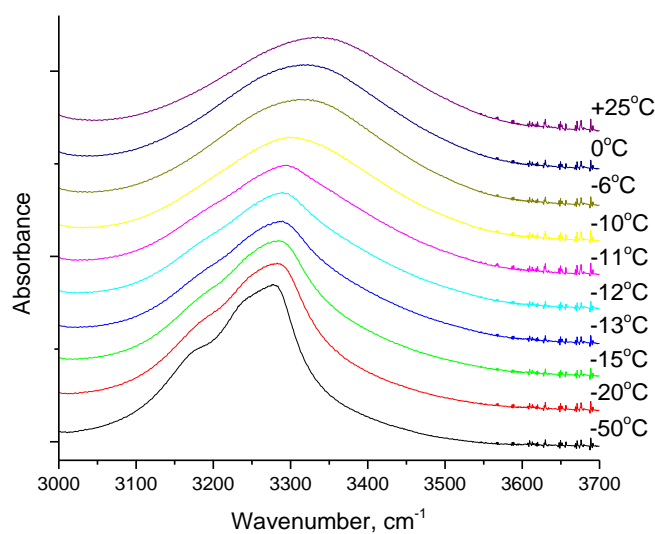

Fig. 3. Experimental FTIR spectra of 1-nonanol for different temperatures in the spectral region 3000 – 3700 cm$^{-1}$

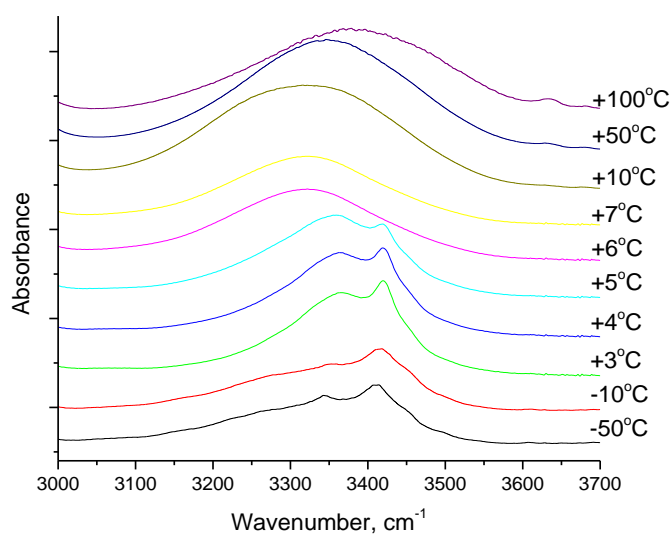

Fig. 4. Experimental FTIR spectra of 1-decanol for different temperatures in the spectral region 3000 – 3700 cm$^{-1}$

Moreover, low frequency edge of the stretching O-H vibration band in 1-decanol become smoother while heating from -50 to 0°C. The same changes can be seen in 1-nonanol while heating from -50 to -10°C.

In fig. 5 and 6, temperature-induced evolutions of the registered FTIR spectra of 1-nonanol and 1-decanol in the spectral region 900 – 1200 cm$^{-1}$ are presented. This region also can be used for interpretation of structural changes in alcohols. For example, one can observe disappearing of bands at 915, 965, 1015 and 1120 cm$^{-1}$ in 1-nonanol. Moreover, bands in the region from 1025 to 1100 cm$^{-1}$ lose their structured form; the band at 1045 cm$^{-1}$ has significant absorbance redistribution to lower frequency region (1035 cm$^{-1}$) and a peak near 1060 cm$^{-1}$ shifts toward 1055 cm$^{-1}$.

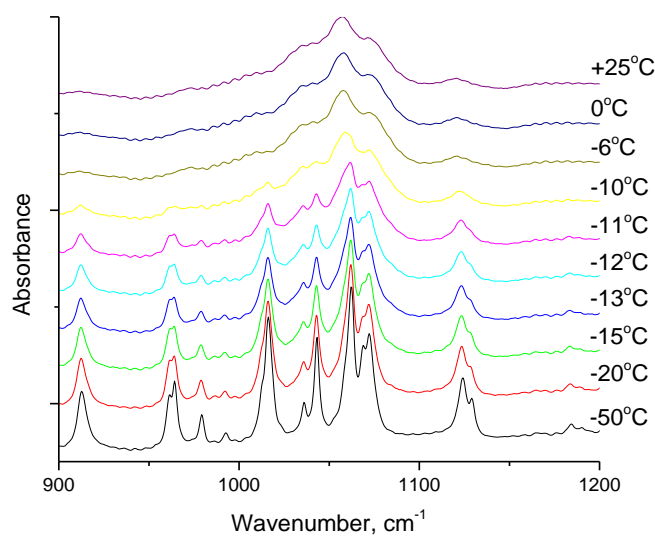

Fig. 5. Experimental FTIR spectra of 1-nonanol for different temperatures in the spectral region 900 – 1200 cm$^{-1}$

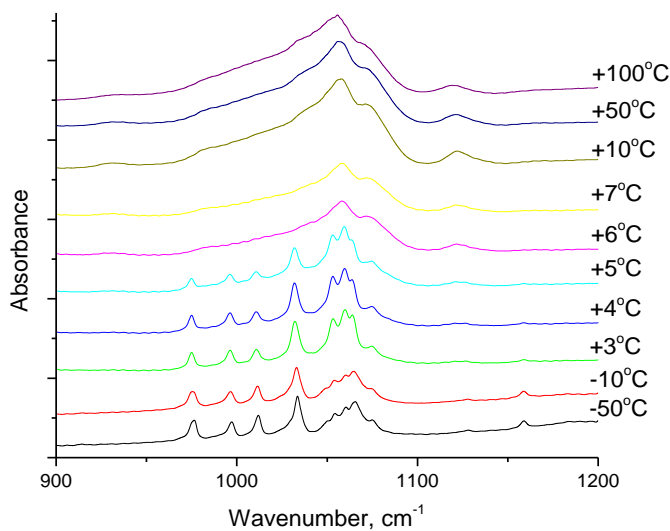

Fig. 6. Experimental FTIR spectra of 1-decanol for different temperatures in the spectral region 900 – 1200 cm$^{-1}$

On the other hand, in 1-decanol we can observe not only disappearing of band at 1160 cm$^{-1}$ and shifts of bands at 975 and 1065 cm$^{-1}$ toward 980 and 1055 cm$^{-1}$ respectively, but also the transformation of structured bands at 995, 1010 and 1032 cm$^{-1}$ into the smooth edge of a wide band with a peak at 1055 cm$^{-1}$. Moreover, we can observe the appearance of new intensive band at 1120 and a weak band at 930 cm$^{-1}$.

In fig. 7 and 8, we can observe more structural changes in the spectral region 1200 – 1600 cm$^{-1}$. Unfortunately, this region cannot be simply used for interpretation of structural changes in alcohols due to the large number of vibrational bands that can easily cover each other. In the spectra of 1-nonanol, we can see more red shifts: from 1465 cm$^{-1}$ to 1457 cm$^{-1}$ and from 1480 cm$^{-1}$ to 1470 cm$^{-1}$. Intensity redistribution and frequency shifts in the spectral region 1340 – 1420 cm$^{-1}$ can reflect a significant change of cluster structure. The same can be seen in the region 1450 – 1480 cm$^{-1}$. According to the literature data, the band at 1450 cm$^{-1}$ is assigned to symmetric bending vibration of C-H bond.

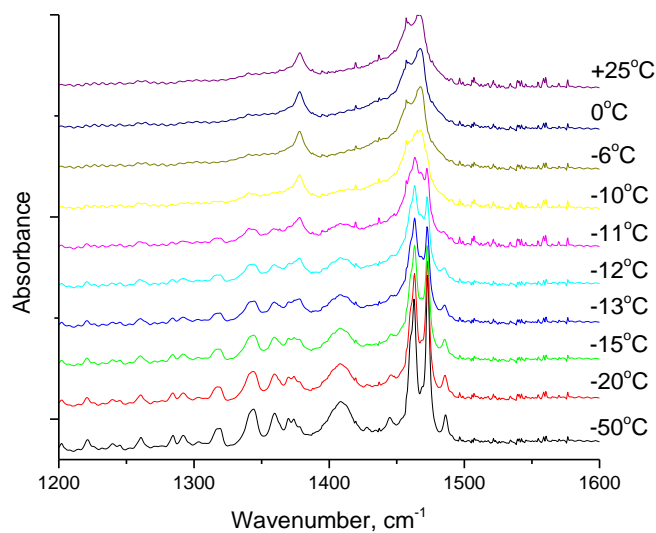

Fig. 7. Experimental FTIR spectra of 1-nonanol for different temperatures in the spectral region 1200 – 1600 cm$^{-1}$

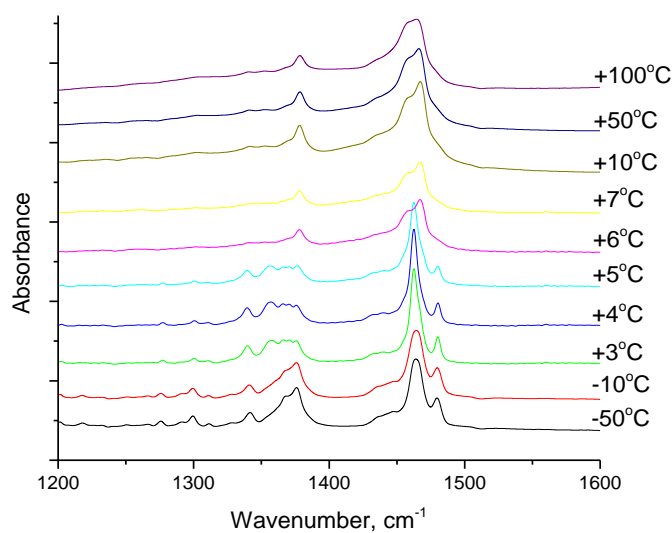

Fig. 8. Experimental FTIR spectra of 1-decanol for different temperatures in the spectral region 1200 – 1600 cm$^{-1}$

In 1-decanol, we also can observe disappearing of bands at 1275 – 1365 cm$^{-1}$. On the other hand, bands at 1375 cm$^{-1}$ and 1465 cm$^{-1}$ shift toward 1380 cm$^{-1}$ and 1470 cm$^{-1}$. These two bands and the band at 1480 cm$^{-1}$ become much weaker. Moreover, new band at 1455 cm$^{-1}$ appears in process of phase transition solid-liquid.

## 4. Comparison

As has been shown in the previous section, spectral changes in process of heating and phase transition in 1-nonanol and 1-decanol are different in every spectral region.

In spectral regions from 900 to 1200 cm$^{-1}$ and from 1200 to 1600 cm$^{-1}$ we observe red shifts in spectra of 1-nonanol and blue shifts in spectra of 1-decanol. Moreover, after the phase transition to liquid, we have seen disappearance of bands in 1-nonanol and appearance of new ones in 1-decanol.

The reason of the observed red shifts is a bigger mass of oscillators per each vibrational band. It is a result of cluster formation: strong external O-H bonds pick over much mass from C-O and other internal bonds. These red shifts mean a redistribution in cluster sizes - from big clusters to small ones. Blue shift means redistribution from smaller to bigger ones.

In addition, the whole spectra in these regions at temperatures lower than the temperature of phase transition for 1-decanol are less structured.

In the region of the stretching O-H vibrations in 1-nonanol, the blue shift is observed while heating, that means redistribution of cluster structures from bigger to smaller ones. Moreover, the spectra of both alcohols do not contain bands at 3600 – 3700 cm$^{-1}$ corresponding to free O-H vibrations (which is present in chain-like clusters or in isolated molecules).

On the other hand, in 1-decanol we can observe a red shift after the phase transition and then – a blue shift while heating from 10 to 100 °C. In addition, there are three different bands in the region 3600 – 3700 cm$^{-1}$. One of them at 3610 cm$^{-1}$ exists up to the phase transition and disappear after structure transformation. Two other bands at 3630 and 3680 cm$^{-1}$ appear simultaneously with a red shift in region 3150 – 3500 cm$^{-1}$. Increasing of intensities at 3600-3700 cm$^{-1}$ and a blue shift of the wide O-H vibrational band could be a result of cluster redistribution from cyclic to chain-like clusters of the same size.

**5. Conclusion**

The temperature-inducted evolution of FTIR spectra of 1-nonanol and 1-decanol was observed. According to the obtained spectra, a comparison of spectral changes in process of phase transformation of 1-nonanol and 1-decanol were made. Our comparison has shown that in phase transformation solid-liquid these alcohols change their cluster structures differently. While 1-nonanol undergo red shifts at the region of 900 – 1600 cm$^{-1}$ and a blue shift at 3000 – 3700 cm$^{-1}$ that correspond to redistribution from bigger clusters in solid state to smaller ones in liquid phase. 1-decanol has blue shifts (900 – 1600 cm$^{-1}$) and red shift (3000 – 3700 cm$^{-1}$), that correspond to transition from smaller to bigger clusters. Moreover, in liquid phase 1-decanol undergo redistribution from cyclic cluster with no free O-H band to chain-like or more complicated clusters with free O-H bonds.